# Magnetoelectric Effect and Phase Transitions in CuO in External Magnetic Fields

**Zhaosheng Wang[1], Navid Qureshi[2], Shadi Yasin[1,3], Alexander Mukhin[4], Eric Ressouche[5], Sergei Zherlitsyn[1,\*], Yurii Skourski[1,\*], Julian Geshev[6,7], Vsevolod Ivanov[4], Marin Gospodinov[8] & Vassil Skumryev[7,9,\*]**

\*Corresponding authors:

Yurii Skourski
Hochfeld-Magnetlabor Dresden (HLD-EMFL), Helmholtz-Zentrum Dresden-Rossendorf, D-01314 Dresden, Germany
Phone: +4935126023421, e-mail: i.scurschii@hzdr.de

Sergei Zherlitsyn
Hochfeld-Magnetlabor Dresden (HLD-EMFL), Helmholtz-Zentrum Dresden-Rossendorf, D-01314 Dresden, Germany
Phone: +493512602617, e-mail: s.zherlitsyn@hzdr.de

Vassil Skumryev
Institució Catalana de Recerca i Estudis Avançats, E-08010 Barcelona, Spain, and
Departament de Física, Universitat Autònoma de Barcelona, 08193 Bellaterra, Barcelona, Spain.
Phone: +34935814872; e-mail: vassil.skumryev@uab.es

[1]Hochfeld-Magnetlabor Dresden (HLD-EMFL), Helmholtz-Zentrum Dresden-Rossendorf, D-01314 Dresden, Germany . [2]Institut Laue Langevin, 6 rue Jules Horowitz, Boîte Postale 156, F-38042 Grenoble Cedex 9, France. [3]American University of the Middle East (AUM), College of Engineering and Technology, Egaila,Kuwait. [4]Prokhorov General Physics Institute, Russian Academy of Sciences, ul. Vavilova 38, Moscow 119991, Russia. [5]SPSMS, UMR-E CEA/UJF-Grenoble 1, INAC, F-38054 Grenoble, France. [6]Instituto de Física, Universidade Federal do Rio Grande do Sul, Porto Alegre, 91501-970 Rio Grande do Sul, Brazil. [7]Departament de Física, Universitat Autònoma de Barcelona, 08193 Bellaterra, Barcelona, Spain. [8]Institute of Solid State Physics, Bulgarian Academy of Sciences, 1784 Sofia, Bulgaria. [9]Institució Catalana de Recerca i Estudis Avançats, E-08010 Barcelona, Spain.

Abstract

Apart from being so far the only known binary multiferroic compounds, CuO has a much higher transition temperature into the multiferroic state, 230K, than any other known material in which the electric polarization is induced by spontaneous magnetic order, typically lower than 100K. Although the magnetically-induced ferroelectricity of CuO is firmly established, no magnetoelectric effect has been observed so far, as direct crosstalk between bulk magnetization and electric polarization counterparts. Here we demonstrate that high magnetic fields of ≈50T are able to suppress the helical modulation of the spins in the multiferroic phase and dramatically affect the electric polarization. Furthermore, just below the spontaneous transition from commensurate (paraelectric) to incommensurate (ferroelectric) structures at 213K, even modest magnetic fields induce a transition into the incommensurate structure and then suppresses it at higher field. Thus remarkable hidden magnetoelectric features are uncovered establishing CuO as prototype multiferroic with abundance of competitive magnetic interactions.

As a building block of high-temperature superconductors, CuO (tenorite) has been extensively studied following their discovery. Unlike other 3$d$-metal monoxides, CuO has a monoclinic crystal structure[1] described by the centrosymmetric space group $C2/c$. Magnetic order sets in at the Néel temperature $T_{N2} \approx 230$ K and the magnetic moments order[2-8] in an incommensurate helix with an envelope oblique to the propagation vector **q** (oblique helix), the so-called AF2 phase. Another transition to a collinear commensurate antiferromagnetic structure, AF1, with the moments along the monoclinic $b$ axis, takes place at $T_{N1} \approx 213$ K.

Several findings, namely multiferroicity[9] of the AF2 phase, electromagnons[10], and presence of toroidal moments[11], have triggered a renewed interest in CuO. Furthermore, it was recently suggested[12] that room-temperature multiferroicity could be expected upon application of hydrostatic pressure. It is worth noting that in some hexaferrites an electric polarization was observed[13] even at room temperatures but only in magnetic field, i.e., the polarization was field induced, not spontaneous as in CuO.

Although the magnetically-induced ferroelectricity of CuO is firmly established, no magnetoelectric effect, i.e., magnetic-field-induced electric polarization changes, has been observed, prompting to call CuO a material with persistent multiferroicity without magnetoelectric effects[14]. If a magnetic field could alter the oblique helix of the magnetic moments in the multiferroic phase (as this happens in the magnetically-driven low-temperature multiferroics) then a strong magnetoelectric effect is expected. We recall[15-17] that in CuO an electric polarization, $P$, may originate from the cycloidal component of the oblique helical structure projected on the plane containing $\mathbf{q}$ and the $b$ axis. It onsets perpendicular to the radial vector $\mathbf{r}_{ij}$ that connects two magnetic moments and is perpendicular to the spin-chirality vector, $(\mathbf{S}_i \times \mathbf{S}_j)$, with $\mathbf{S}_i$ and $\mathbf{S}_j$ being the moments of those atoms: $\mathbf{P} \sim \mathbf{r}_{ij} \times (\mathbf{S}_i \times \mathbf{S}_j)$. The apparent correlation between $P$ and the long-pitched magnetic structure is often evidenced by magnetic-field control of $P$. In the case of CuO, so far it has only been demonstrated that an electric field could partly change the population of magnetic domains with opposite chirality[18] and that a dynamic magnetoelectric coupling exists at THz frequencies resulting in electromagnons[10].

Tentative magnetoelectric phase diagrams of CuO have been recently suggested by the analysis of a nonlocal Landau-type free energy from symmetry arguments[19,20] or by use of the praphase concept and exchange symmetry[21]. Those models are backed by only few experimental results, namely bulk magnetization and ultrasound velocity at relatively low magnetic fields. The latter data suggest that a third phase, AF3, with presumably incommensurate sinusoidal magnetic structure originally proposed in Ref. 22, appears at $T_{N3}$. It exists in a narrow temperature range of less than 1 degree right above $T_{N2}$ (Ref. 18), as also indicated by recent thermal-expansion measurements[23]. Such phase was not reproduced in the earlier theoretical studies[24-26]. Firmly establishing the magnetoelectric phase diagram of CuO will help to understand the physics of the broad family of monoclinic multiferroic systems. It would also eventually point to magnetoelectric effects at relatively high temperatures.

Here the magnetoelectric phase diagram of this high-temperature magnetically-induced multiferroic is sketched, unveiling additional phase transitions and important magnetoelectric features. We demonstrate that sufficiently high magnetic fields of ≈ 50 T are able to suppress the helical modulation of the magnetic moments in the multiferroic phase and dramatically affect the electric polarization. Furthermore, just below the spontaneous magnetic transition from commensurate (paraelectric) to incommensurate (ferroelectric) magnetic structures at 213 K, even modest magnetic fields induce a transition into the incommensurate magnetic structure and then suppresses it at higher field, thus causing remarkable polarization changes.

## Results

**Zero-magnetic-field magnetic structures and transitions.** The results of our magnetic susceptibility, magnetic torque, magnetocapacitance, thermal expansion, ultrasound velocity and single crystal neutron diffraction experiments (Supplementary Notes 1 and 2, Supplementary Figures 1, 2 and 3a) are in a general agreement with the known sequence of the temperature-induced magnetic transitions at zero magnetic field[2-9,18,27,28], i.e., successive AF1 → AF2 → paramagnetic state transitions. The refined magnetic structures are visualized in Figs.1 a,b,d.

**Magnetic-field-induced transitions.** Magnetic-field-induced transition into a high-field (HF1) phase takes place when the magnetic field **H** is applied along the *b* axis within the collinear AF1 phase below 213 K[19,29]. The transition is believed to be of a spin-flop type with the magnetic moments flopping perpendicular to **H**, thus causing a jump of the magnetization at the critical field $H_{cr}$. As seen in the magnetization curves (Fig. 2a), $H_{cr}$ ≈ 10.5 T at 2 K. The field-induced transition is reflected in some of the neutron diffraction reflections (Supplementary Figure 3b). We have refined the magnetic structure, unknown before, of this HF1 phase at 12 T. It consists of collinearly aligned moments [0.51(1)$\mu_B$] within the *ac* plane at an angle of 26(2)° from the *c* axis towards the positive *a* axis (Fig. 1c). Note that, importantly, this direction is close to the second axis of the helical envelope, i.e., the collinear magnetic moments flop from one envelope axis to the other, perpendicular to **H**.

As expected, $H_{cr}$ increases with $T \to T_{N1} \approx 213$ K, i.e., towards the commensurate – incommensurate transition temperature at $H = 0$ (Fig. 2a). Surprisingly, we found that $H_{cr}$ suddenly decreases at the vicinity of $T_{N1}$, as manifested in the magnetization curve at 212 K. The magnetic-field evolution of the neutron diffraction rocking curves at 212 K (Fig. 2c) reveals a transition from the non-polar collinear AF1 to an incommensurate magnetic structure with $q$ = (0.509 0.000 −0.483) as in the polar AF2 phase, with the two magnetic phases coexisting in a range of fields around 10.5 T as implied by the overlap of the respective intensities. This observation strongly suggests that an electric polarization could eventually appear under application of modest magnetic fields at temperatures just below $T_{N1}$, which would constitute evidence for a magnetoelectric effect. To verify this suggestion, integrated magnetic-peak intensities were collected at 211.4 K. The zero-field diffraction data can be explained by the collinear low-temperature model, i.e., the magnetic moments [0.29(1)$\mu_B$ at this temperature] are collinearly aligned along the $b$ axis. At this temperature, a magnetic field of 12 T is just enough to induce a transition into the incommensurate oblique helical phase with $q$ = (0.509 0.000 −0.483) [first envelope axis: 0.22(2)$\mu_B$ along the $b$ axis; second envelope axis: 0.34(4)$\mu_B$ at an angle of 20(6)° from the $c$ axis], i.e., into the AF2 phase. We note that the high-field data within the incommensurate phase could be explained equally well by a collinear model with the moments within the $ac$ plane (as the HF1 structure). However, the symmetry of this magnetic model is not compatible with the observed macroscopic electric polarization described below. The limited neutron diffraction data set, due to the aperture of the cryomagnet, namely lacking magnetic Bragg reflections with non-zero $k$-indices, is the reason why we cannot unambiguously determine the complex magnetic structure by the performed neutron diffraction alone. It is worth noting that the ultrasound velocity measurements (Fig. 2b and also Supplementary Note 3 and Supplementary Figure 4), reveal that the magnetic-field-induced transition for $T < 211.5$ K is actually reflected in two ultrasound anomalies, a result which we attribute to the fact that in a real sample the spin-flop occurs via an intermediate state existing within few kOe, as our magnetization curves show (Fig. 2a), and softening of the sound velocity at two fields of magnetic instability: for the low-field AF1 and for the high-field HF1 phases, correspondingly. Also, note that this two-stage behaviour disappears for $T > 211.5\,\text{K}$ but lower that $T_{N1}$, which is consistent with AF1 to AF2 field-induced transition taking place in this temperature range.

In search for a field-induced transition within the multiferroic incommensurate AF2 phase (from 213 to 230 K, approximately), the magnetization at 218 K has been recorded in pulsed magnetic fields as high as 60 T applied along the *b* axis as well as along the direction $\alpha$ within the *ac* plane, i.e., along two perpendicular axes of the helical envelope. For those two directions, one would expect a sharp change of magnetization at a critical field for which the plane of the magnetic helix is reoriented in a direction close to orthogonal with respect to **H**. This would result in switching **P** from *b* to *a* or/and *c* axes. However, note that other high-field spin configurations including collinear structures cannot be ruled out. The latter scenario would lead to *P* = 0.in the high-field phase. The magnetization measured in both directions fails to provide unambiguous proof for any field-induced transitions. This, however, could be a consequence of the proximity of $T_N$ and the specific characteristics of the zero-field helical structure, both of them influencing the sharpness of the expected spin-flop transition. We also note (Supplementary Figure 5) that the magnetization at 60 T is ≈ 0.07 $\mu_B$ f.u.$^{-1}$(formula unit), much smaller than that expected for a spin-flip "ferromagnetic" alignment of spin 1/2 $Cu^{2+}$ (forced by **H**) with a moment of 1 $\mu_B$, thus reflecting the strength of the exchange coupling interactions.

**Magnetic-field effect on electric polarization.** Next, we turn to the other counterpart, electric polarization, **P**, which is known to exist at zero-magnetic field in the AF2 phase and is directed along the monoclinic *b* axis. As a direct probe of polarization changes, the pyrocurrent was measured in pulsed magnetic field **H**$_b$ applied along *b* axis. Figure 3a shows that at ≈ 215.5 K, a temperature which is well inside the AF2 phase, the pyrocurrent $I_b$ measured along *b* reveals either a maximum or a minimum at $H \approx 37\,\text{T}$, depending on the sign of the poling electric field *E* (~ 170 V mm$^{-1}$) in which the sample was cooled from $T > T_{N2} \approx 230\,\text{K}$ (or alternatively warmed up from $T < T_{N1} \approx 213\,\text{K}$) to ensure a predominant orientation of the polar domains. Such behaviour constitutes a clear fingerprint of a magnetoelectric effect within the incommensurate phase, implying in turn a field-induced magnetic transition and consequent polarization change. A tiny hump was recorded on $I_b$ at $H \approx 37$ T, for the descending branches of the $H$-pulse. Note that no $E$ was present during the $H$-pulse in the above measurement. The tiny change of $I_b$ on reducing $H$ should be attributed to only small misbalance in the number of magnetic domains with opposite chirality when the AF2 phase is re-entered without $E$ applied (see also Supplementary Note 4). In fact, when an electric field equal to the poling $E$ holds during the $H$-pulse (Supplementary Figure 6a), the changes of $I_b$ for the two branches of the magnetic field sweep are similar in magnitude (since the balance between the magnetic domain with left and right chirality is impaired on re-entering the AF2 phase under $E$) and opposite in sign. The small difference in the positions of the maximum and the minimum of $I_b$ could be attributed to a hysteresis in the field-induced transition (presumably of first-order type) and/or to dynamic effects[30].

A special case among the $I_b(H_b)$ is the one measured at temperatures just below $T_{N1} \approx 213$ K, where no polarization should exist at H = 0, e.g., at 212.5 K (Fig. 3b). Unlike $I_b(H_b)$ at temperatures between $T_{N1}$ and $T_{N2}$, where only one minimum and one maximum is observed for each of the ascending and descending branches of the $H$ sweep, at 212.5 K a maximum and minimum are seen for the ascending branch followed by a maximum and minimum for the descending branch. We recall that at 212 K we observed a field-induced AF1→AF2 transition at $H_{cr} \approx 10$ T (Fig. 2). This critical field is in a good agreement with the lower field-induced transition manifested on the $I_b(H_b)$ data. Thus we conclude that the two transitions manifested on each of the $I_b(H_b)$ branches are related to consecutive crossings of the borders between AF1→AF2→HF1 with $P_b = 0$ phase on increasing $H$ (and in reversed order, on decreasing $H$). This finding constitutes a solid proof for the magnetoelectric effect in CuO and particularly for the remarkable magnetic-field-induced appearance and disappearance of $P$ at a fixed temperature within a narrow temperature region below $T_{N1} \approx 213$ K.

The electric polarization estimated by integrating the pyrocurrent data with respect to time is plotted in Fig. 3 at some selected temperatures; the complete data set is given in

Supplementary Figure 6. Because of the high coercivity of *P*, rather high poling fields are needed to establish a single-domain state[18,31], which should be the reason for the slight underestimation of *P* as compared to the highest literature values.

**Ultrasound velocity and magnetostriction.** To further probe the magnetic field-induced transitions, we used magnetostriction and ultrasound techniques, both sensitive to spin-lattice coupling (Fig. 4).

The temperature dependence of the sound-velocity change $\Delta v/v$ of the transverse ultrasonic wave propagating along *b* (**k** || **H** || *b* and **u** || *c*) measured for **H** || *b* (Fig. 4a) reveals anomalies matching the known spontaneous transitions and those induced by *H* as discussed above (see also Fig. 2b). The temperature hysteresis on the data below 213 K (the left bottom inset) reflects the first-order nature of the AF1→AF2 transition in presence and in absence of **H**. Importantly, the ultrasound velocity data provide strong support to the recently promoted claim[19-21,23] that an additional AF3 magnetic phase exists, sandwiched between the paramagnetic and the AF2 phases (top right inset) as well as for the evolution of its phase borders in presence of magnetic field. Because of the very narrow temperature interval of existence (just ≈0.2 K at H = 0) and mainly because of its proximity to the paramagnetic state, our attempts to characterize this phase by neutron diffraction did not succeed. It is plausible to assume, by analogy with other monoclinic multiferroics (e.g., $MnWO_4$), that the AF3 phase has an incommensurate sinusoidal collinear structure with a spin direction along *b*.

The magnetostriction at 212.4 K and for H = 0 (Fig. 4b) exhibits anomalies at the critical fields of ≈12 and ≈ 17.5 T in good agreement with the field-induced AF1→AF2 and AF2→HF1 transitions, respectively, as found in our magnetization and in the neutron diffraction experiments. Similarly, the critical field of ≈ 34 T at 215.2 K corresponds to the transition from AF2 to HF1 phase. We notice that the *b* axis shrinks (since $\Delta L<0$) at the field-induced AF1→AF2 transition taking place at ≈ 12 T at 212.4 K. This agrees with the thermal expansion at the same spontaneous transition (Supplementary Figure 7). At the same time, *b* expands ($\Delta L>0$) upon the field-induced AF2→HF1 transition at ≈17.5 T for 212.4 K and at ≈ 34 T for 215.2 K. We also note that the magnitude of the changes at those transitions is rather similar.

Within the experimental resolution of the pulsed-field experiment, we were not able to detect magnetostriction changes at the AF1→HF1 spin-flop transition at low temperatures. However, more sensitive measurements in steady $H$ show (Supplementary Figure 8) that a minute positive $\Delta L$ change takes place at this transition implying diminutive $b$ axis expansion in the HF1 phase. At $T$ higher than ≈ 219 K, no peculiarity has been detected within the resolution of any of our magnetostriction experiments. From the fact that $\Delta L$ is finite between the incommensurate AF2 and the two commensurate AF1 and HF1 phases but is much smaller between the two commensurate ones, we may draw the conclusion that the absence of measurable magnetostriction change at the transition into the high-temperature high-field phase favours an incommensurate phase. Considering the phase diagrams of well-investigated multiferroics like $MnWO_4$ or $Ni_3V_2O_8$, it is plausible to assume that this HF3 phase has a sinusoidal modulation with the moments perpendicular to $b$.

**Magnetoelectric phase diagrams.** Figure 5a displays the phase diagram for $H \parallel b$ constructed on the base of the critical fields recorded on bulk magnetization, pyroelectric current, magnetostriction, ultrasound, and neutron diffraction data. Magnetocapacitance data (Supplementary Figure 9) are also plotted. We note that only tiny changes in the pyrocurrent in the other two crystallographic directions, $I_a(H_b)$ and $I_c(H_b)$, were observed (probably due to spurious currents or/and such originating from the polar domain walls) implying that **P** does not switch direction but vanishes upon the **H**-driven transformation of AF2. Besides, we have sketched in the phase diagram the hypothetical[20] phase boundaries (dashed lines) between the sinusoidal collinear phases (along $b$ and perpendicular to it) and the paramagnetic phase.

The pyrocurrent $I_b$ was also measured for **H** applied along the distinct direction $\alpha$ within the $ac$ plane, which is the other axis of the helical envelope (Fig. 3b). Unlike $I_b(H_b)$, $I_b(H_\alpha)$ is zero below $T_{N1} = 213$ K, while for $T$ between $T_{N1}$ and $T_{N2}$, $I_b$ has a similar trend and a single minimum/maximum for the ascending/descending branches of $H$, respectively. Again, only tiny changes in the pyrocurrent in the other two crystallographic directions, $I_a(H_\alpha)$ and $I_c(H_\alpha)$, were observed (probably due to spurious currents) implying that **P** does not switch direction but vanishes upon the **H**-driven transformation of AF2. The phase diagram constructed on the base of the critical fields recorded on the pyroelectric current together with the hypothetical phase boundaries is given in Fig. 5b.

**Discussion**

In what follows we will attempt to justify the constructed phase diagrams. According to the antiferromagnetic resonance data in the low temperature collinear phase of CuO[32,33], the resonance frequencies of two modes corresponding to spin oscillations closer to the *bc* and *ba* planes (strictly speaking, these planes should be the easy $\alpha b$ and hard $\alpha' b$ planes where $\alpha' = \alpha + 90^\circ$) are $\nu_1 = \gamma(2H_{A1}H_E + H_{A1}^2)^{1/2} \approx 308$ GHz and $\nu_2 = \gamma(2H_{A2}H_E + H_{A2}^2)^{1/2} \approx 1852$ GHz, respectively, where $H_E$ is the exchange field and $H_{A1}$ and $H_{A2}$ are the anisotropy fields in the corresponding planes ($H_{A1,A2} << H_E$). This implies that the *ba* plane's anisotropy energy exceeds that in the *bc* plane by $\approx 36$ times since $(\nu_2/\nu_1)^2 = (2H_{A2}H_E + H_{A2}^2)/(2H_{A1}H_E + H_{A1}^2) \approx H_{A2}/H_{A1}$. It allows considering CuO as an "easy plane" antiferromagnet, in which the helical phase can be stabilized only in the easy *bc* plane. At zero magnetic fields, the development of the magnetic structure is governed by the fine balance between the exchange and the anisotropy energies. Just below the Néel temperature, the magnetic structure is determined by the exchange energy favouring an incommensurate structure with spins along an easy *b* axis, i.e., a sinusoidal phase (AF3 promoted in Refs. 19-22, whose existence is supported by our work, see the top right inset in Fig. 4a).However, in this structure the gain in exchange energy is not full because of a strong reduction of the average local spins (magnetization) at some crystallographic sites owing to the collinear spin arrangement and the sinusoidal modulation of the moment size. In order to increase the gain in the exchange energy, the helical AF2 structure in an easy plane formed by the $\alpha b$ spin rotation plane is developed in spite of some increase in the anisotropy energy. However, when the temperature is further lowered, this increase of the anisotropy energy becomes unfavourable and, in order to minimize the total energy, the system again goes to a collinear spin configuration along *b* (AF1) though in a commensurate antiferromagnetic state (losing in exchange energy but gaining in anisotropy energy).

For **H** || *b* one could expect a spin flop of the cycloid to the *ac* plane. However the strong anisotropy related with a spin declination towards the hard *a* axis (more precisely, to the hard axis $\alpha'$ close to *a*) does not favour such *ac* plane flopped cycloid state. Instead, a field-induced transition into HF1 (as shown by our neutron data) or HF3 phases takes place (both having collinear structures with moments perpendicular to *b*). None of these two flopped magnetic structures supports polar order. Our data imply the existence of a triple point in the high-field part of the phase diagram at which the AF2-HF1, the AF2-HF3 and the HF1-HF3 phase boundaries meet as well as the absence of the HF2 phase, i.e., the supposedly flopped helix.

Similarly, for **H** || $\alpha$ within the spin-rotation plane one could expect an annihilation of the $\alpha$ component resulting in a commensurate or incommensurate sinusoidal modulation of magnetic moments along *b*.

Thus, it seems that the role of the strong anisotropy of CuO discussed above has been underestimated in Refs. 19 and 20, where a detailed phase diagram has been simulated for **H** || *b*. In the light of the above arguments, the predicted spin-flop helical phase HF2, which supports polar order, should be absent and instead a direct transition between HF1 and HF3 phases should take place.

To conclude, magnetoelectric effect was observed in CuO - a prominent simple oxide with distinct position among the multiferroic materials, as suppression of the spontaneous electric polarization at high magnetic fields. Furthermore, it was found that just below the spontaneous magnetic transition from commensurate (paraelectric) to incommensurate (ferroelectric) magnetic structures at 213 K, even modest magnetic field induces a transition into the incommensurate magnetic structure and then suppresses it at higher fields, thus causing remarkable polarization changes at rather high temperatures. This utilizes a possibility of abruptly-induced electric polarization by low magnetic fields. The magnetoelectric phase diagrams of this high-temperature magnetically-induced multiferroic have been sketched, unveiling unknown transitions and important magnetoelectric features.

**Methods**

**Crystal growth:** Single crystals of CuO were grown by the high temperature solution growth method. The starting materials (minimum purity of 99.999%) were CuO powder and flux of 95% $Bi_2O_3$ and 5% $B_2O_3$ in ratio of 20:1. The mixture was sealed in a platinum crucible, maintained at 1150 °C for 48 h and then slowly cooled to 920°C at a rate of 0.5 degrees per hour. At 920 °C the flux was removed from the crucible by decanting. The crystal quality was checked by neutron back scattering. A sample of size 1.5×1.5×1.5 mm³ was cut along the principle crystallographic directions and then used in all the experiments. The resistivity of our crystal shows highly isolating character (about 50 MΩ mm along the *b* axis at room temperature) suggesting nearly stoichiometric ratio of Cu and O.

**Pulsed magnetic fields:** The facilities at the Dresden High Magnetic Field Laboratory (HLD-EMFL) in Dresden were used. The experiments in this study were performed in pulsed magnetic fields up to 60 T with pulse duration of 35 and 150 ms.

**Magnetization and torque measurements:** Magnetization and magnetic torque in DC fields up to 14 T were measured using commercial SQUID magnetometers from Quantum Design. None of the data have been corrected for ionic diamagnetic core contributions or for demagnetizing-field effect since those corrections are negligibly small. Pulsed-field magnetization was measured by integrating the voltage induced in a coil system surrounding the sample. The system was precisely compensated prior to the pulse in order to cancel the d*B*/d*t* contribution.

**Ultrasound measurements:** Ultrasound measurements in steady and pulsed magnetic fields were performed using a pulse echo technique. Two piezoelectric lithium niobate ($LiNbO_3$) resonance transducers were glued to opposite parallel surfaces in order to excite and detect acoustic waves. Transverse acoustic waves with wave vector **k** were propagated along the *b* axis with polarization **u** directed along *c* axis. The sound velocity *v* and the sound attenuation Δ*α* were measured at an applied magnetic field parallel to the *b* axis. The ultrasound frequency was between 33 and 111 MHz.

**Polarization, magnetocapacitance and magnetostriction:** Dielectric polarization *P* was measured by a pyroelectric technique[34]. The pyrocurrent was captured through the voltage variation in a shunt resistor connected in series with the measurement circuit by a digital oscilloscope Yokogawa DL750 with a high sampling rate of 1 MS/s and a resolution of 16 bit. Then *P* was calculated by integrating the pyrocurrent numerically.

The capacitance is measured by a high precision capacitance bridge GR1615-A. The excitation signal and the voltage of the bridge circuit are recorded by the digital oscilloscope Yokogawa DL750. After the pulse, the signal processing is performed by a lock-in procedure simulated in the computer.

The optical fiber Bragg grating (FBG) method[35] was used to measure magnetostriction in both steady field up to 18 T and in pulsed magnetic fields up to 60 T. The magnetostriction along the *b* axis of the crystal was measured with the magnetic field applied along *b*. The relative length change Δ*L*/*L* can be obtained from the shift of the Bragg wavelength of the FBG.

**Neutron diffraction:** The magnetic structures at zero-magnetic field have been investigated on the D23 diffractometer (CEA CRG ILL, Grenoble) in combination with a 4-circle cryostat, as well as on the D10 diffractometer at ILL in order to have a reference point for the subsequent experiments with a magnetic field applied parallel to the crystallographic *b* axis. Roughly 500 unique magnetic reflections have been measured at selected temperatures within the commensurate phase and within the incommensurate phase. Symmetry analysis has been applied in order to deduce those magnetic moment configurations which are compatible with the nuclear symmetry. The magnetic structures under an applied field along the *b* axis have been analysed with a vertical cryomagnet at D23. Due to the limited aperture of the cryomagnet, 50 unique magnetic reflections have been collected.

**Acknowledgements**

The authors are grateful to Virginie Simonet, Michel Viret, Michael Kuz'min, Josep Fontcuberta and Florencio Sanchez for critical reading of the manuscript. We would like to acknowledge the support from the Institute Laue–Langevin (ILL – Grenoble) for the neutron diffraction experiments and the support from HLD at HZDR, a member of the European Magnetic Field Laboratory (EMFL), for the high magnetic field experiments. The work was supported by the Bulgarian Science Fund FNI-T-02/26 project, by the French ANR "Dymage" project ANR-13-BS04-0013, by the Spanish Government (MAT2011-29269-C03 and MAT2014-56063-C2-1-Rprojects), by the Generalitat de Catalunya project 2014 SGR 734, by the Brazilian agencies CNPq (project 407033/2013-0) and CAPES (BEX 0298/2015-08), and by the Russian Foundation for Basic Researches project 13-02-01093.

**Author contributions**

M.G. grew the single crystals. Z.W. carried out the electric polarization, magnetostriction and magnetocapacitance experiments and data analysis. E.R and N.Q. carried out the single crystal neutron diffraction experiments and refined the magnetic structures. S.Z. and S.Y. performed the ultrasound velocity experiments and data analysis. J.G. and V.I. contributed to data analysis and interpretation. Y.S. provided the pulse field magnetization data. V.S. initiated the project and participated in all the experiments. A.M. and V.S. wrote the manuscript with assistance from N.Q. and input from E.R. and S.Z. All authors contributed to the scientific discussion and edited the manuscript.

**Additional information**

Supplementary information is available in the online version of the paper. Reprints and permissions information is available online.

**Figure Captions:**

**Figure 1 | Visualization of the magnetic structures investigated by single-crystal neutron diffraction.**

(**a**) The incommensurate antiferromagnetic oblique helical structure (AF2) at 215 K, where the magnetic moments rotate in a plane passing across *b* axis and making an angle of 74(2)° with the propagation vector $q$ = (0.5060.000–0.483). See also the d-panel. (**b**) Commensurate antiferromagnetic structure (AF1) at 10K with chains of alternating magnetic moments pointing along the *b* axis. (**c**) High-field spin-flopped phase (HF1) with the magnetic moments pointing at 26(2)° from the *c* axis within the *ac* plane. (**d**) Visualization of the projected cycloid. The envelope of the circular helix (depicted in red with its main axes) makes an angle of 74(2)° with the propagation vector. Its projection onto the plane defined by the propagation vector and the crystallographic *b* axis corresponds to the cycloidal component of the spin rotation with envelope and main axes shown in green.

**Figure 2 | Magnetic-field-induced transitions in CuO.** (**a**) Magnetization curves at different temperatures. For temperatures below the transition into the AF1 phase at $T_{N1}$ = 213 K, jumps of magnetization are observed at $H_{cr}$ for the magnetic-field-induced transitions; note the reduction of $H_{cr}$ for the 212 K data (see also the inset). (**b**) Magnetic-field dependence of the sound velocity $\Delta v/v$ of the transverse ultrasonic wave with wave vector **k** propagating along the *b* axis with polarization **u** directed along *c* axis (**k** || **H** || *b* and **u** || *c*) measured at different temperatures. Results for up and down field sweeps are shown. The data at different temperatures are shifted arbitrary along the vertical axis for clarity. (**c**) Commensurate to incommensurate field-induced phase transition in CuO at 212 K: neutron rocking curves (rotation of the crystal around the diffractometer vertical axis) as a function of *H*. The position A3 =162.8° corresponds to $q$ = (0.500 0.000 –0.500), whereas $q$ = (0.509 0.000 –0.483) for A3 = 164.5°. The scattered intensity is colour coded between 0 and 2.86x10$^3$ counts per 6s.

**Figure 3 | Magnetic-field dependence of the pyrocurrent and the electric polarization for the two axes of the helical envelope.** (**a**) The pyrocurrent along the *b* axis, $I_b$, as a function of the magnetic field *H* applied along the *b* axis, measured at ≈215.5 K (within the incommensurate phase) after cooling through the Néel temperature in an electric field with different polarity (~170 V mm$^{-1}$). The small shift between the

two sets of data along the *T* axis is a result of small difference (≈ 0.2 K) between the temperatures for the two measurements. The arrows indicate increasing/decreasing *H*. (**b**) $I_b$, as a function of *H* applied along the *b* axis at two selected temperatures (slightly below 213 K and in the incommensurate phase above it). (**c**) $I_b$ as a function of *H* applied along the distinct direction α within the *ac* plane. (**d**) The electric polarization **P$_b$** along the *b* axis at some selected temperatures, as a function of *H* applied along the *b* axis. (**e**) **P$_b$** as a function of *H* applied along the distinct direction α. The data recorded at increasing *H* are presented by solid line, while the data at decreasing *H* are presented by dashed line.

**Figure 4│Magnetostriction effect and magnetic-field effect on the sound velocity for H along *b* axis.**

(**a**) Temperature dependence of the sound velocity Δ*v*/*v* of the transverse ultrasonic wave with wave vector **k** propagating along the *b* axis with polarization **u** directed along *c* axis (**k** || **H** || *b* and **u** || *c*) measured at different applied magnetic fields. Results for up and down temperature sweeps are shown. The (i) inset shows the sound velocity in the vicinity of the phase transitions $T_{N2}$ and $T_{N3}$ after subtracting the linear anharmonic contribution; the (ii) inset shows some details in the sound velocity at the phase transition $T_{N1}$. The arrows indicate the anomalies at the phase transitions. The data at different *H* are shifted arbitrary along the *y*-axis for clarity. (**b**) Magnetostriction (the relative length change Δ*L*/*L*) along the *b* axis recorded in pulsed magnetic field along the same crystallographic axis at two selected temperatures.

**Figure 5│The magnetoelectric phase diagrams of CuO.** (**a**) Magnetic field *H* versus temperature *T* for **H** || *b* axis, based on pyrocurrent (blue dots), capacitance (green dots), magnetostriction (pink dots), sound velocity (red dots) and bulk magnetization data (black dots). Dashed lines are expected hypothetical phase boundaries predicted from theoretical analysis[20]. AF1 stands for collinear commensurate (CM) antiferromagnetic (AFM) structure, with the moments along the *b* axis; HF1 - collinear commensurate antiferromagnetic structure, with the moments perpendicular to the *b* axis; AF2 - hellicoidal, incommensurate (ICM) antiferromagnetic structure; AF3 - incommensurate sinusoidal collinear structure with a spin direction along *b* axis; HF3 - incommensurate sinusoidal collinear structure with a spin direction perpendicular to the *b* axis. $P_b$ is the electric polarization along the *b* axis. (**b**) **H** || α within the *ac* plane (≈30° from the *c* axis) based on pyrocurrent (blue dots) together with the hypothetical phase boundaries (dashed lines).

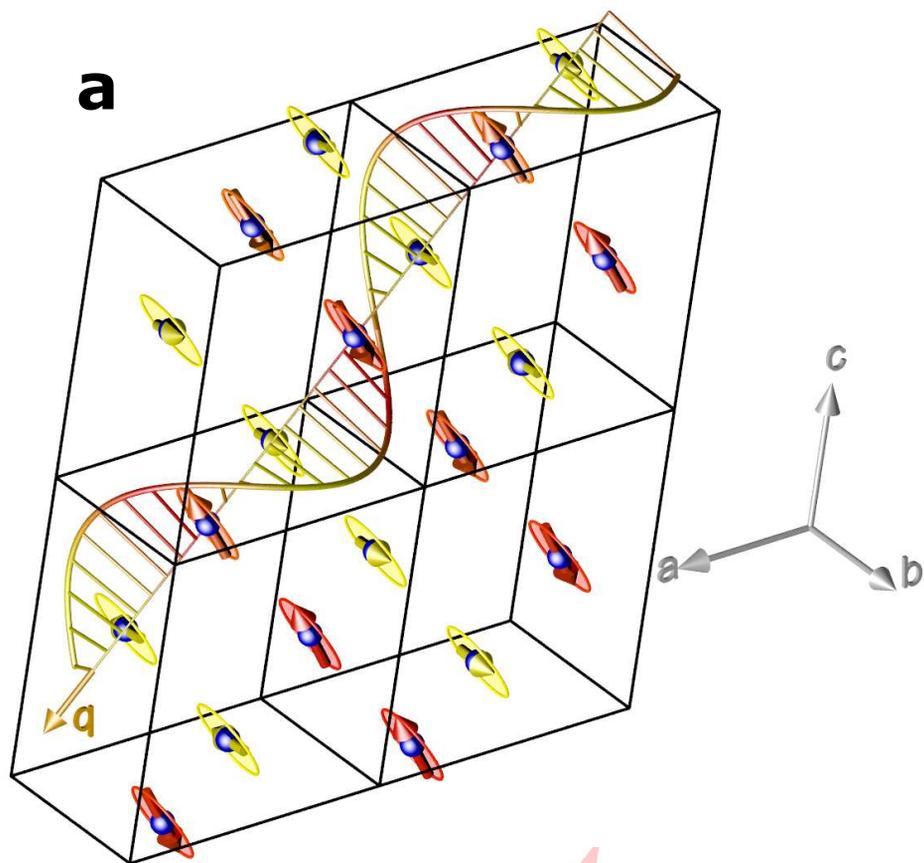
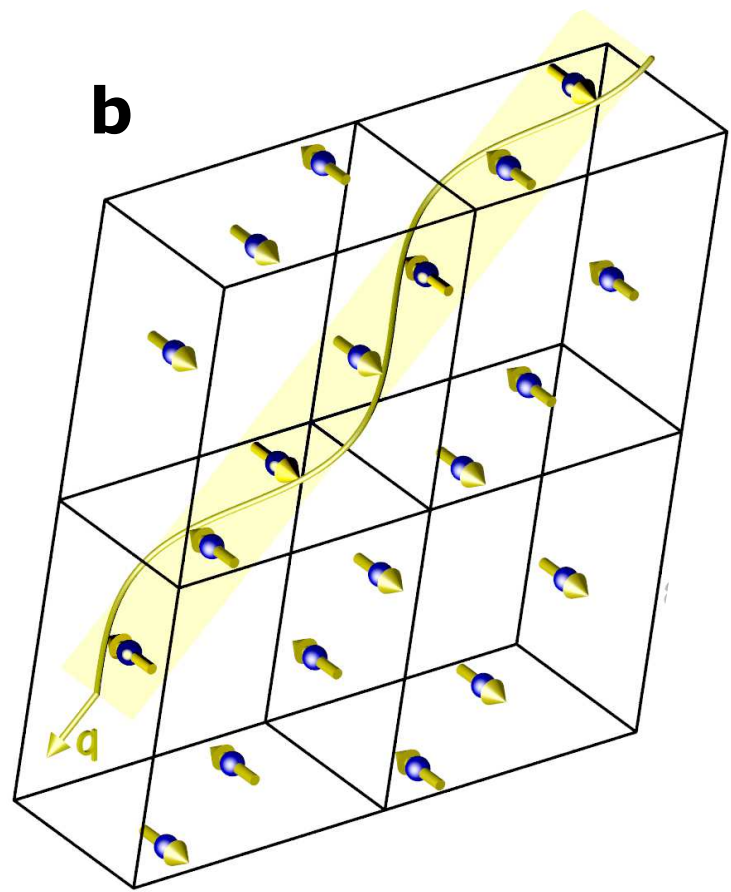
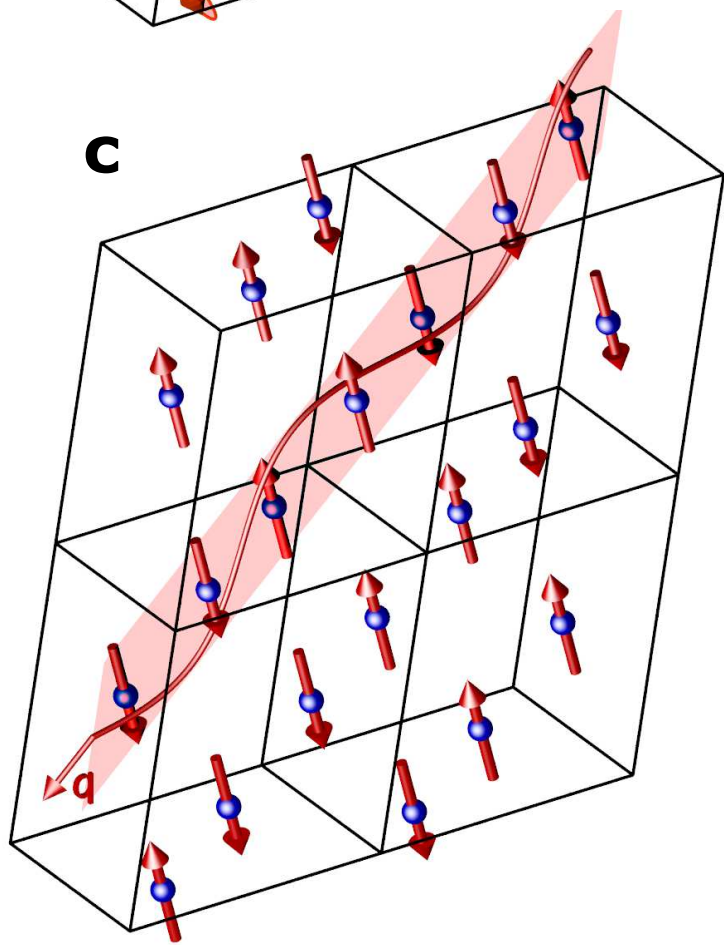
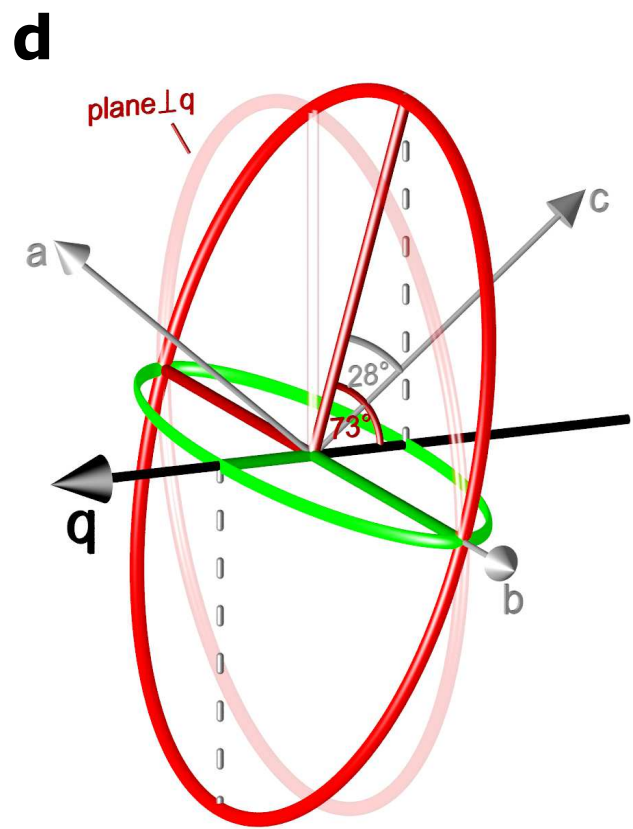

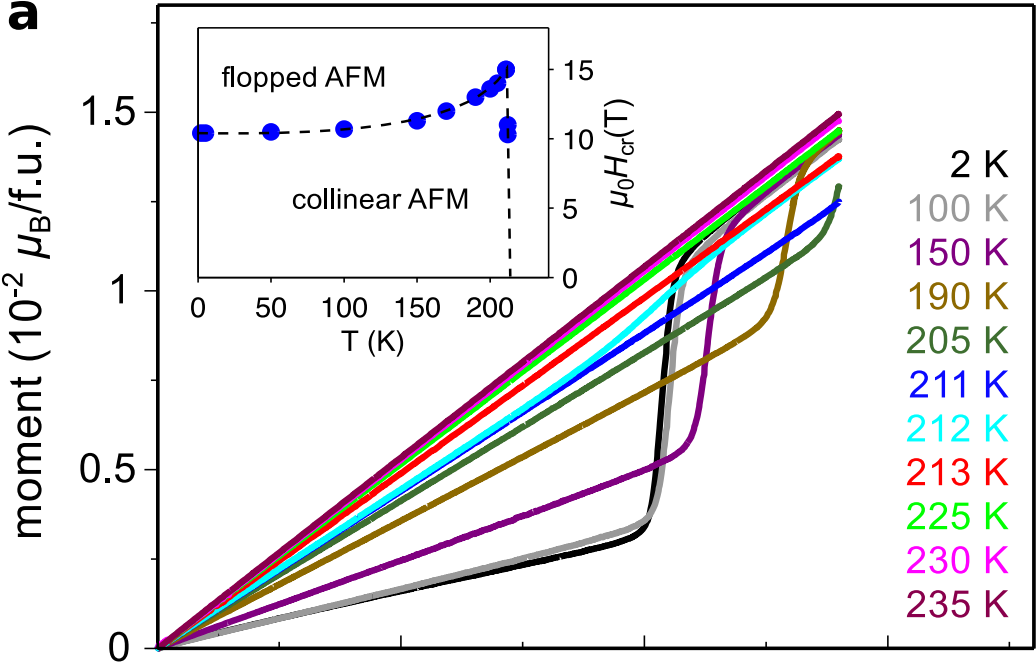
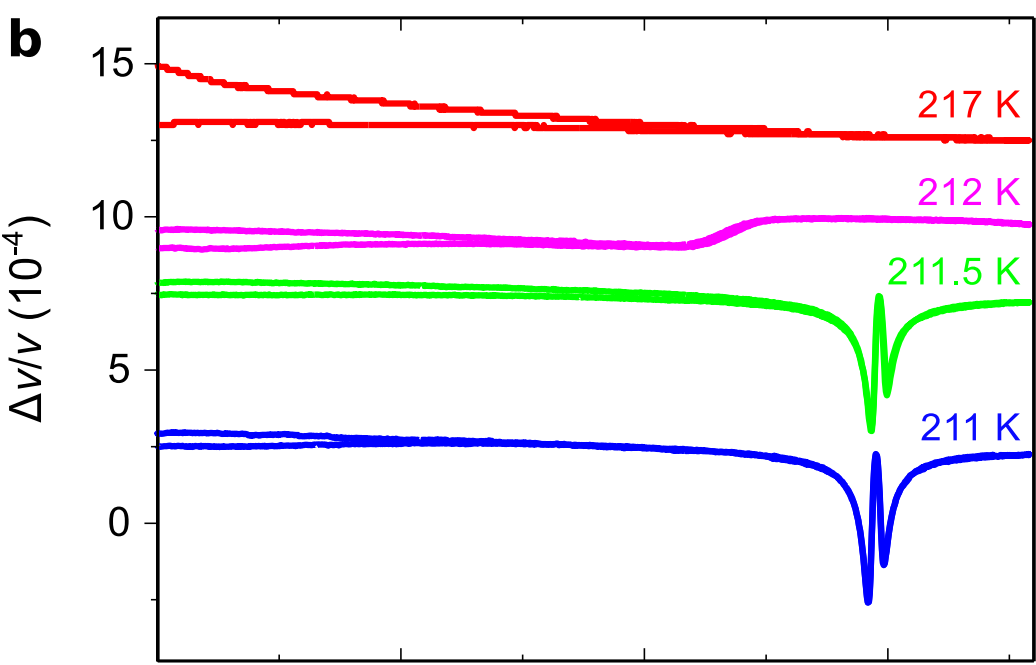
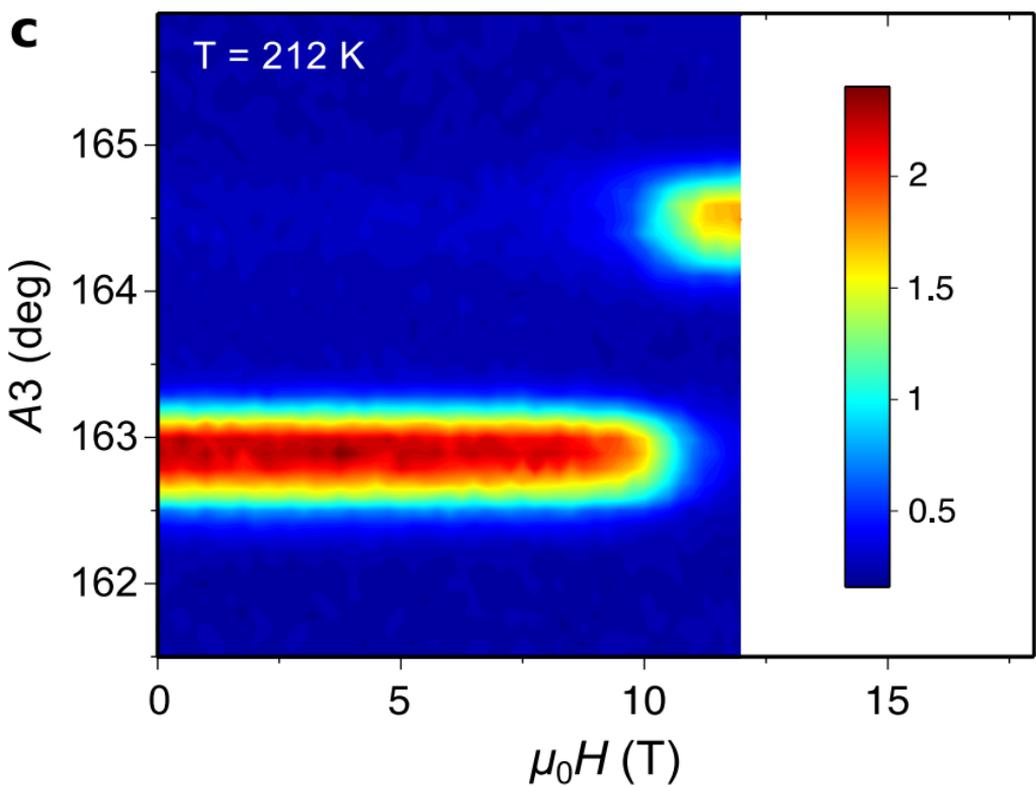

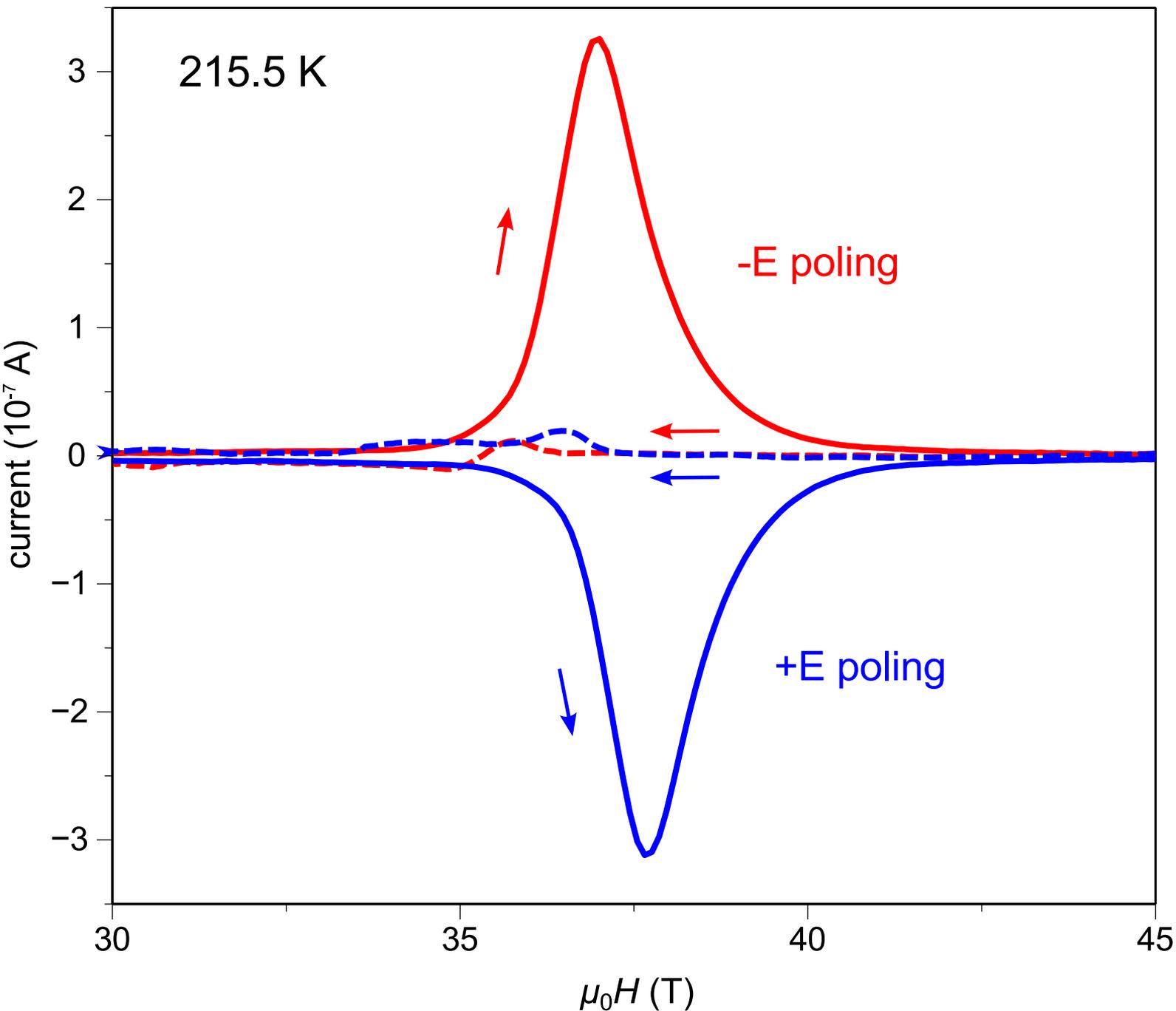

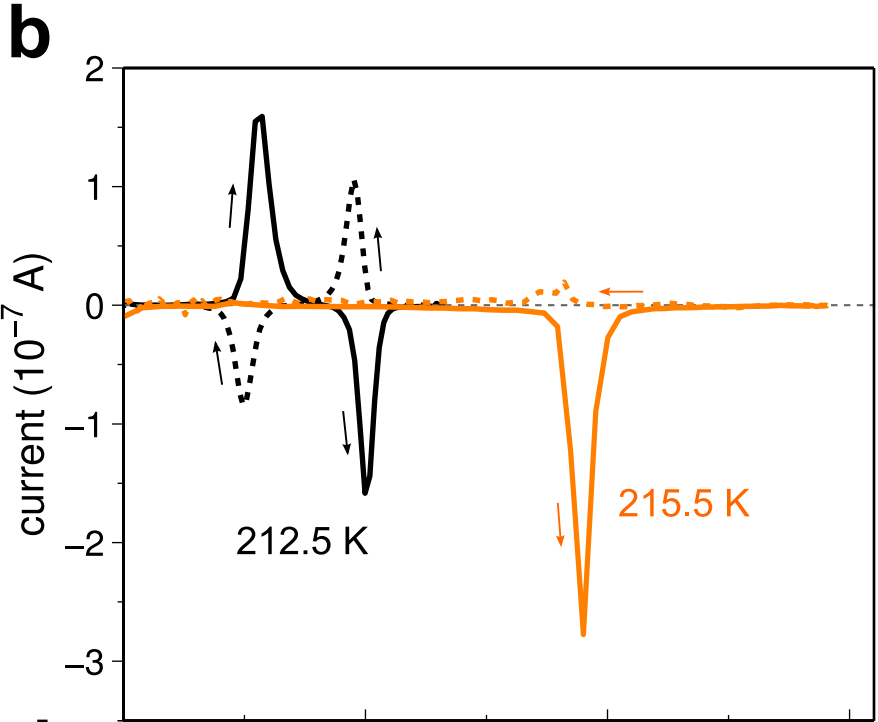
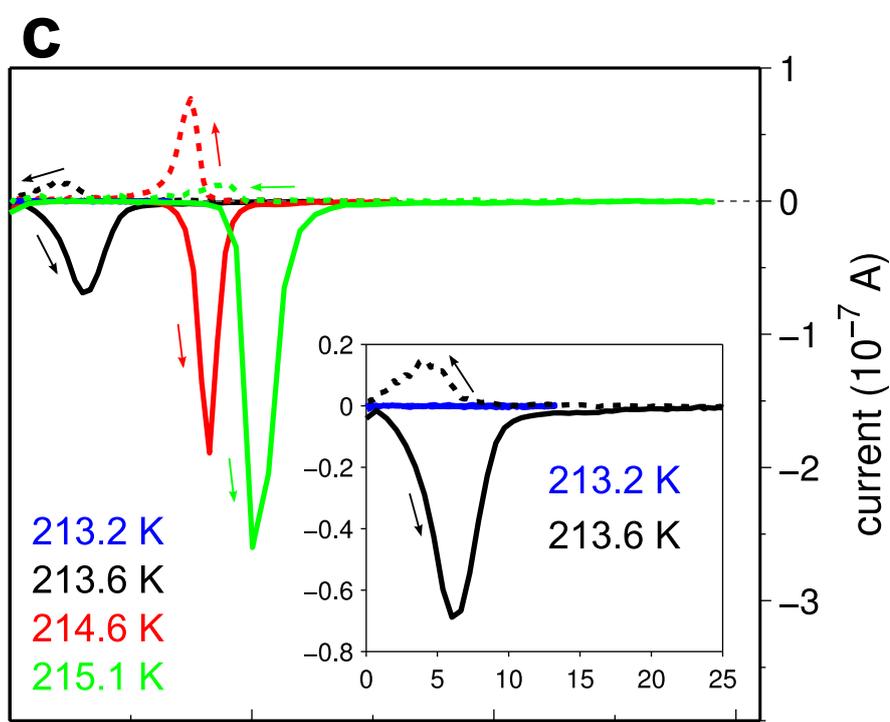
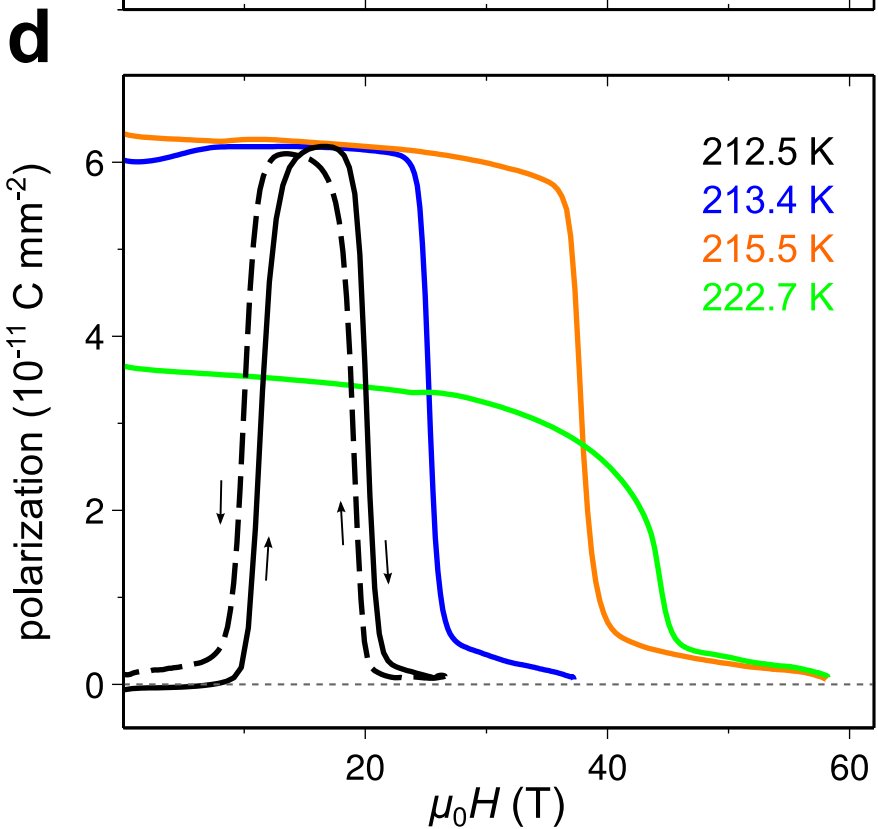
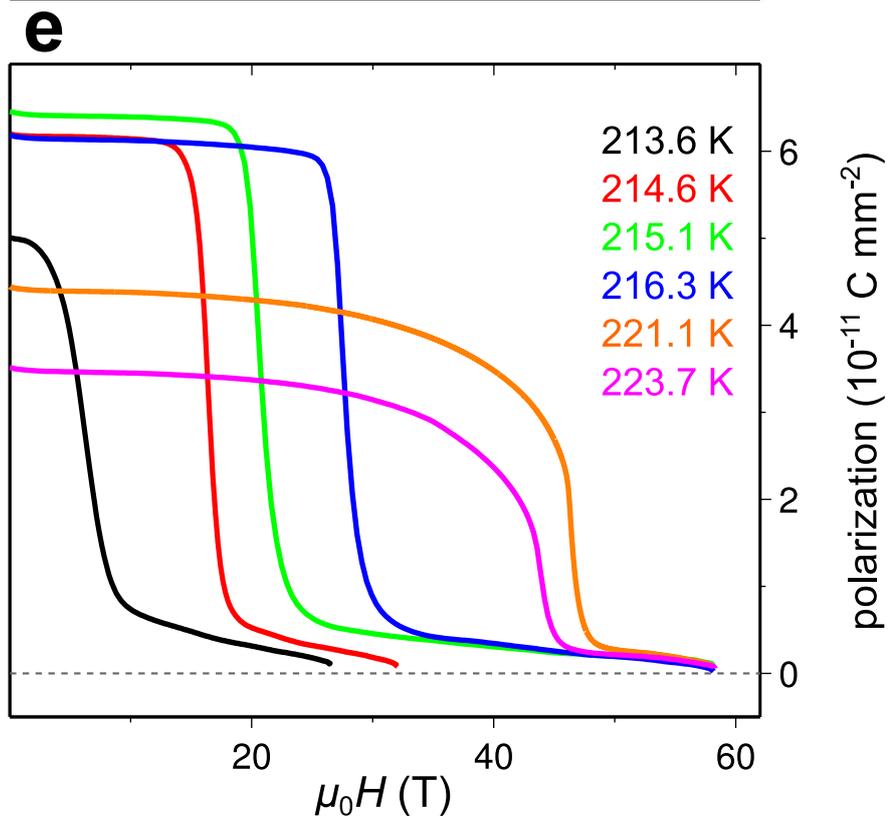

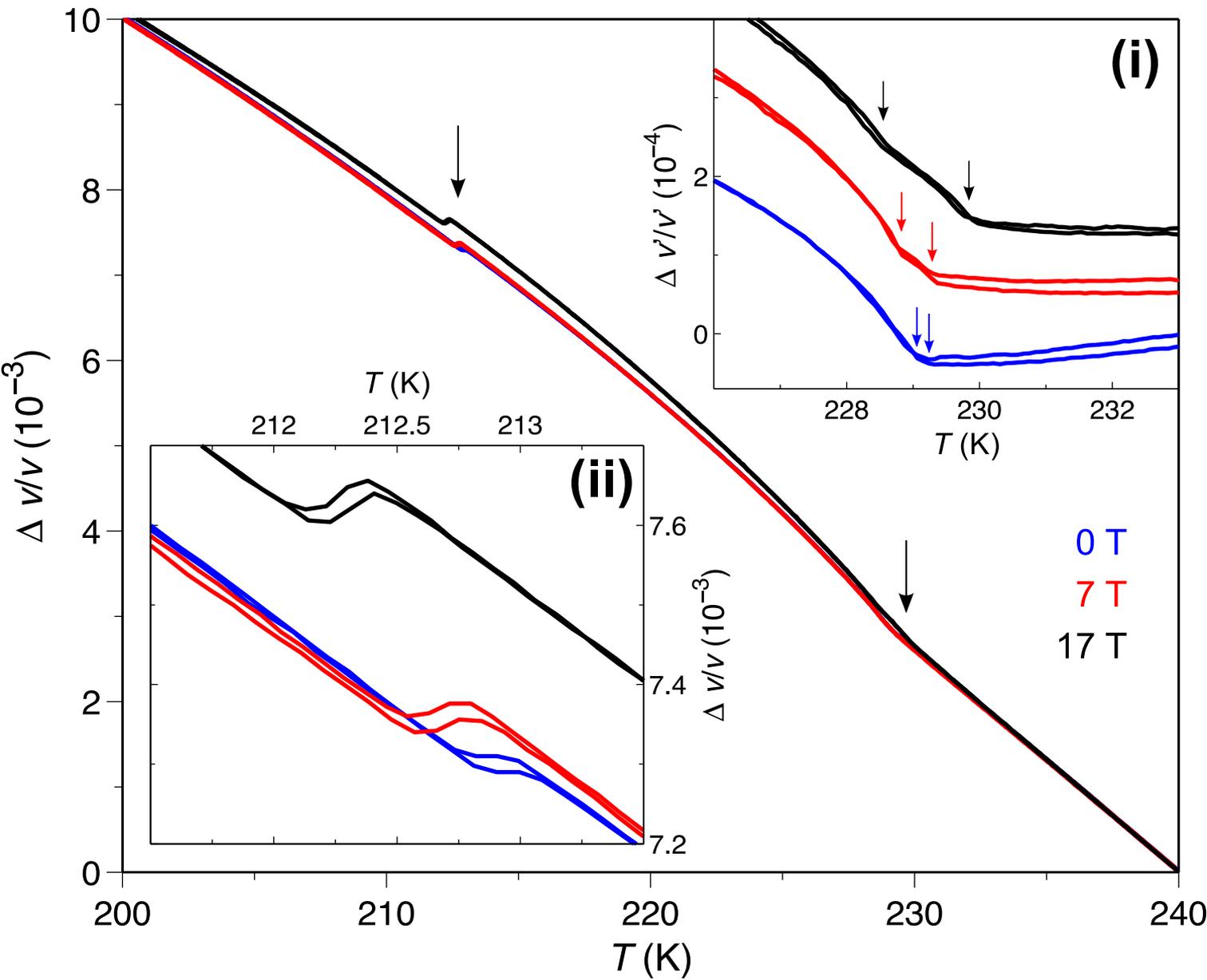

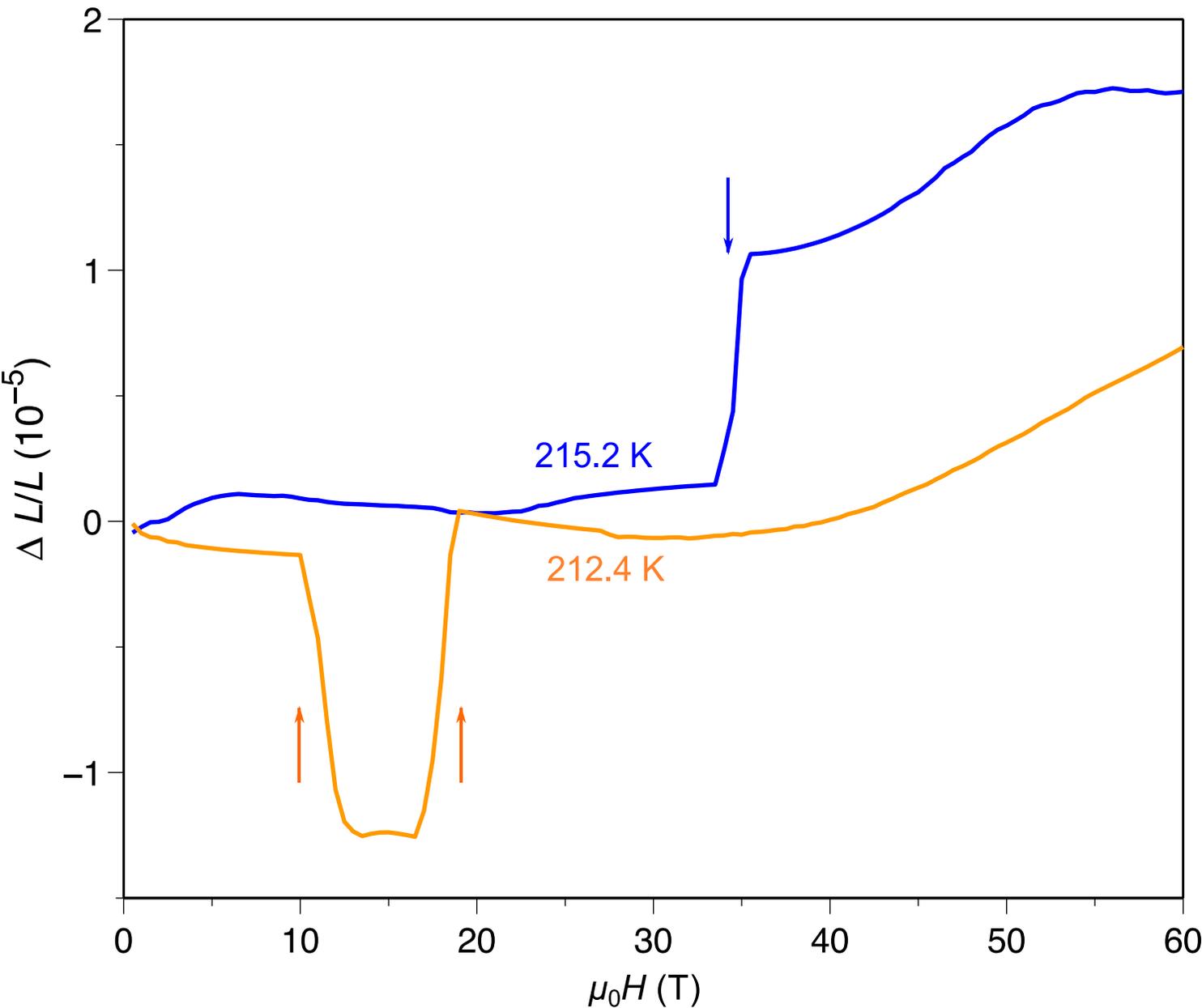

**a**

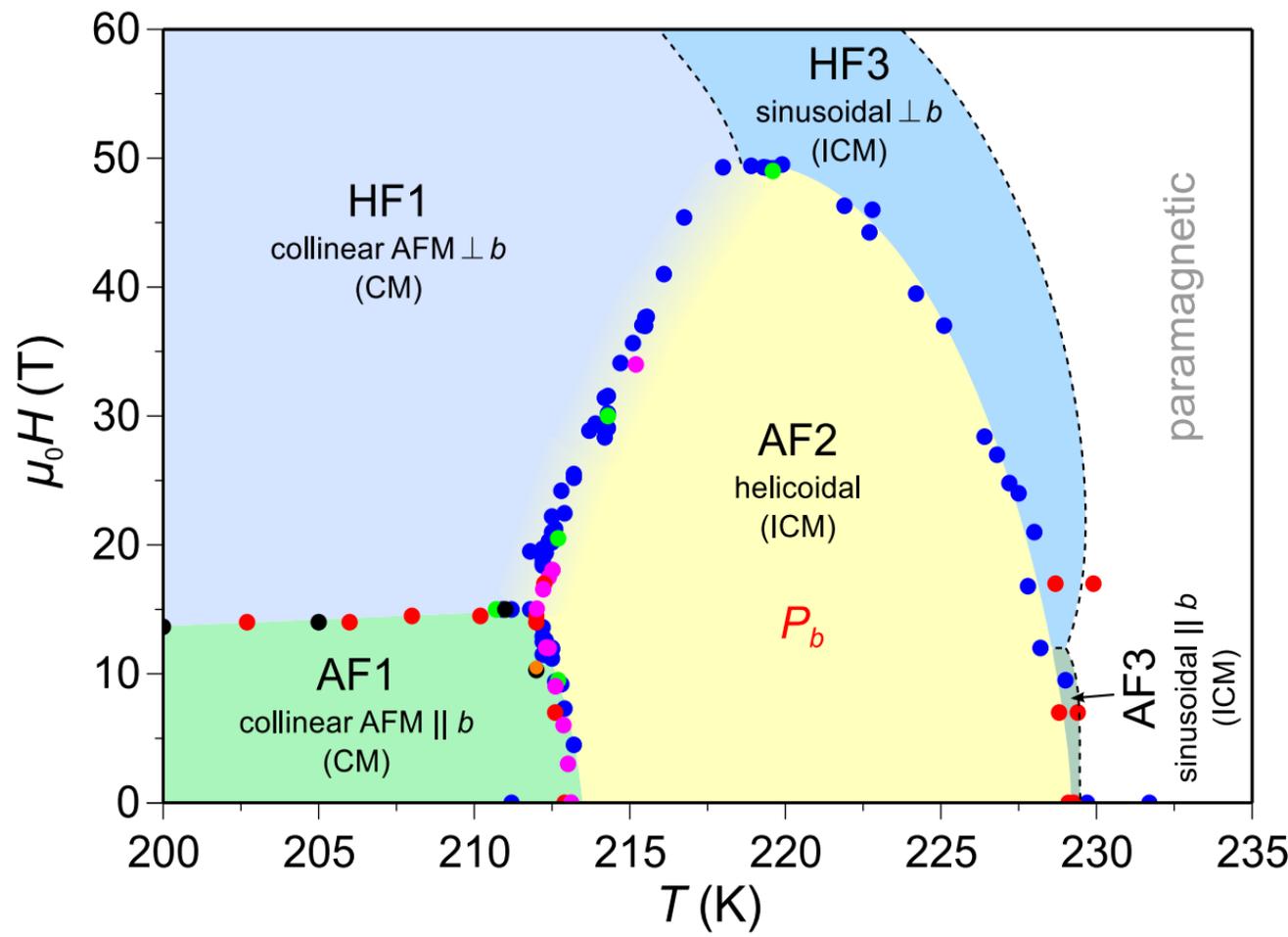

**b**

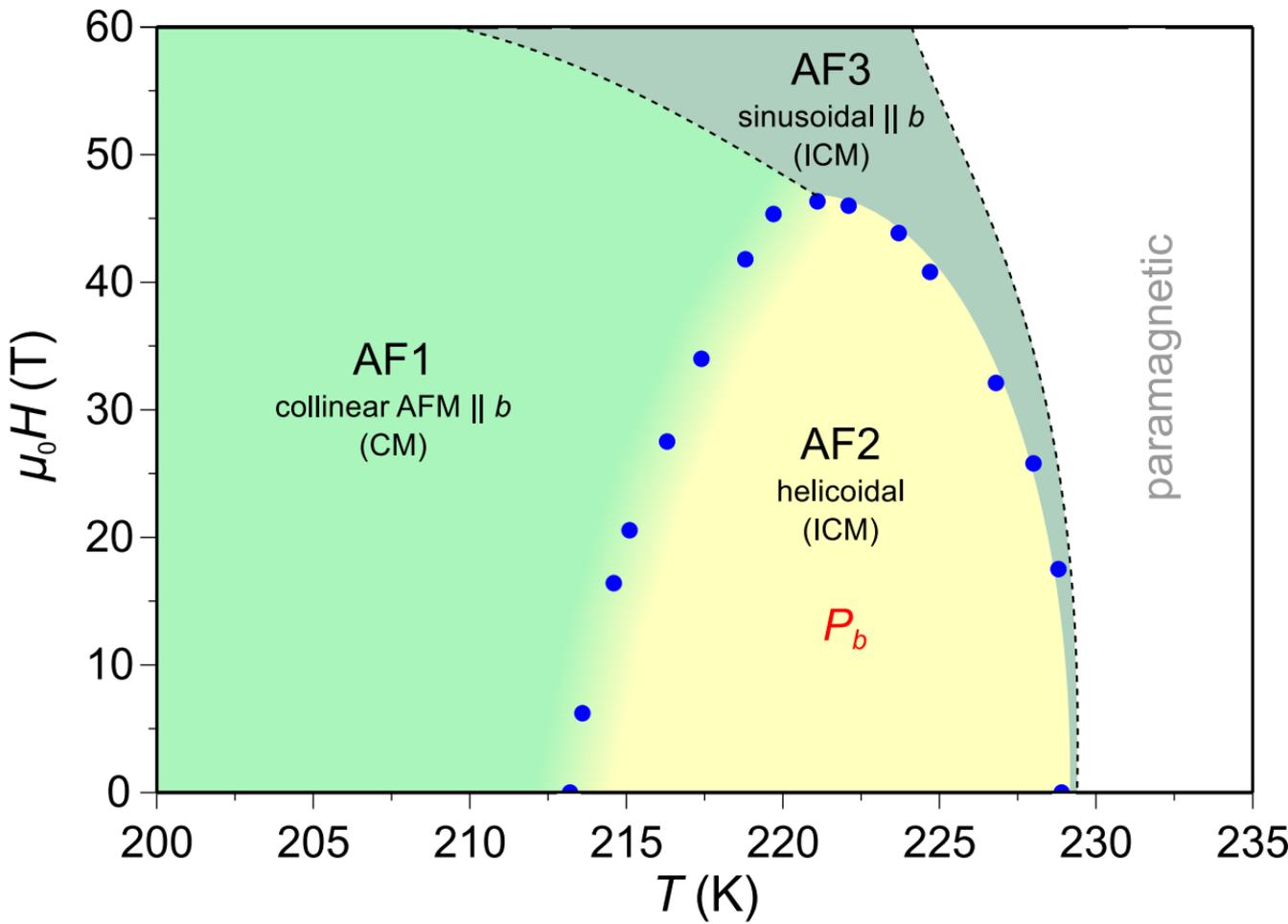